# A Framework for Post Quantum Migration in IoT-Based Healthcare Systems

Asif Alif*, Khondokar Fida Hasan*[†], Basker Palaniswamy[‡], Md. Morshedul Islam[§]
*University of Wyoming, USA
[†]University of New South Wales (UNSW), Canberra, Australia
[‡]University College Cork (UCC), Ireland
[§]Concordia University of Edmonton, Edmonton, AB, Canada
*Corresponding author: k.fidahasan@yahoo.com

*Abstract*—Smart healthcare industry is increasingly relying on Internet of Things (IoT) devices to improve patient care and operational efficiency. However, the cryptographic algorithms that enable fundamental security and are widely used in these cyber systems are vulnerable to attacks by emerging quantum computers - known as Quantum Threat. This paper examines the quantum threat to healthcare IoT across the four layers of the IoT architecture: physical, network, perception, and application. It proposes a comprehensive migration framework integrating a phased hybrid approach with crypto-agility to transition healthcare IoT systems to quantum-safe cryptography. This framework prioritises resource-constrained devices, emphasises interoperability, and considers the challenges of vendor readiness and infrastructure upgrades. This paper contributes a detailed, phased migration plan specifically tailored to the unique security needs and resource limitations of IoT-based healthcare systems.

## I. INTRODUCTION

A major concern of cybersecurity is the rapidly developing field of quantum computing, especially for industries that depend on long-term data integrity and confidentiality. The Healthcare sector is considered one such critical infrastructure that is increasingly dependent on the IoT infrastructure for a wide range of applications, including remote patient monitoring, data management, and operation of sophisticated medical equipment. This dependence on interconnected devices has revolutionised healthcare delivery by enabling real-time patient care, improving diagnostic capabilities, and enhancing operational efficiency. However, this reliance on IoT technologies also expands the attack surface, making healthcare systems more susceptible to cyber threats, particularly in the face of rapidly advancing quantum computing capabilities [1].

Quantum computing presents a formidable challenge to current cryptographic systems that protect cyber infrastructure. Its ability to leverage quantum phenomena like superposition and entanglement, coupled with algorithms like Shor's and Grover's, allows it to efficiently solve mathematical problems exponentially faster, threatening widely used cryptosystems in both asymmetric and symmetric cryptography [2]. This vulnerability arises from the potential of quantum computers equipped with these algorithms to undermine the security protocols and products that rely on these well-known cryptosystems, as illustrated in the symbolic Fig. 1. These risks become amplified in the context of IoT-based healthcare systems, where any compromise in data confidentiality or device integrity could have life-threatening consequences. Hence, it is imperative that healthcare systems begin migrating to quantum-safe cryptographic solutions.

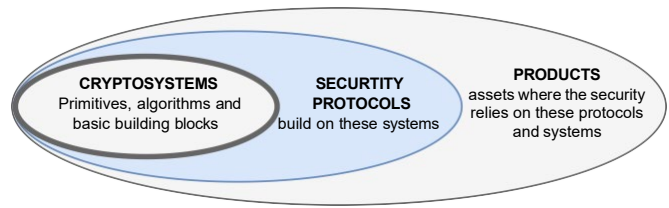

Fig. 1: Assumed vulnerability under Quantum Threat

Despite the urgent need to address the growing quantum threat, integrating post-quantum cryptography (PQC) into IoT-based healthcare systems remains a formidable challenge. Healthcare IoT ecosystems are made up of highly heterogeneous and resource-constrained devices that operate with limited computational, memory, and energy capacities. These limitations significantly hinder the adoption of computationally intensive PQC algorithms. The architectural structure of IoT systems varies across the literature, with some researchers advocating for a three-layer model, while others propose four or five layers depending on system complexity and security requirements [3]. This study adopts a four-layer IoT architecture- Physical, Network, Perception, and Application- which is widely recognized in academic research for assessing security vulnerabilities in IoT environments [4]. While IEEE P2413 (IoT Architecture Framework) does not explicitly define a layered model, its framework aligns conceptually with the layered approach, particularly in its emphasis on sensing (Perception Layer), communication (Network Layer), and application services (Application Layer) [5]. The inclusion of a Physical Layer in our paper is intended to address critical hardware security concerns, such as tampering and quantum threats targeting embedded systems. Although this specific four-layer structure is not explicitly defined in IEEE or IETF standards, our approach aligns with established security frameworks by incorporating IEEE P2413 for architectural structuring and IETF RFC 8576 (IoT Security Best Prac-tices) for security guidelines, thereby ensuring a robust and comprehensive approach to IoT security [5], [6]. The four layers of IoT architecture - physical, network, perception and application - each present unique vulnerabilities that quantum- enhanced attacks could exploit [4], [7]. Each layer in this architecture is exposed to different classes of quantum threats. For example, quantum brute-force attacks can target cryptographic algorithms in embedded systems at the physical layer, quantum key recovery attacks threaten communication security at the network layer, quantum DDoS attacks compromise the perception layer, and quantum hammer attacks endanger the IoT application layer [4]. These multidimensional vulnerabilities underscore the need for a tailored and strategic migration plan that operates across all layers of the IoT stack, rather than treating the system as a monolithic entity.

While several organisations and researchers have proposed PQC migration strategies for general enterprise or governmental infrastructures, there is currently no comprehensive, structured migration framework specifically tailored for IoT-based healthcare systems. Most existing strategies either generalise the migration approach or overlook the constraints posed by resource-limited devices. This creates a significant gap in the current body of research and highlights the need for a IoT based healthcare sector-specific strategy that ensures both security and operability throughout the transition to quantum-resilient cryptography. This paper addresses this critical gap by proposing a detailed, phased migration framework designed explicitly for IoT-based healthcare systems, taking into account their layered architecture and operational constraints. The proposed framework incorporates cryptographic asset inventorying, IoT-layer-specific quantum risk assessments, algorithm selection based on device constraints, and phased implementation of hybrid cryptographic mechanisms which offers a systematic and secure transition path to PQC, preserving system functionality, enabling interoperability, and enhancing quantum resilience across the IoT based healthcare domain.

## II. QUANTUM THREAT ANALYSIS

Quantum computing constitutes a significant breakthrough in information processing, utilising the fundamental principles of quantum mechanics. It leverages qubits, which, through the phenomenon of superposition, can exist in multiple states that are either 'on' (1) or 'off' (0), at once. When coupled with quantum entanglement, this capability allows quantum computers to execute complex computations at speeds that greatly exceed those of classical computing systems [8]. However, this development also presents a serious threat to established cryptographic techniques, especially those that rely on public-key cryptography, like RSA and ECC. Expert opinion suggests a 1% to 6% chance that quantum computers capable of breaking current cryptographic standards will emerge within the next 5 years [2].

The rapid advancement of quantum computing not only threatens the security of IoT-based devices, which rely on cryptographic algorithms like RSA and ECC to protect data over Wi-Fi, Bluetooth, and Zigbee, but also poses a significant threat to the overall IoT infrastructure. The healthcare sector has emerged as a crucial area of study for cybersecurity because of the quick adoption of IoT technologies, which have revolutionized patient care by enabling advancements in data-driven therapy, monitoring, and diagnostics. This section provides an overview of quantum threats in cryptography and the IoT, with a particular focus on healthcare-related IoT devices.

### A. Quantum Threats to Cryptography

Cryptography, the science underpinning secure communication, is fundamentally based on two types of algorithms: Symmetric Key and Asymmetric Key algorithms. Symmetric Key algorithms, also known as single-key encryption, require the same key to be shared by both the sender and receiver for the processes of encryption and decryption. Conversely, Asymmetric Key algorithms, like RSA, use a public and private key for encryption. The security of RSA relies on the computational challenge of factoring large integers; however, quantum computers, with the right algorithms, can break RSA encryption with relative ease. For example, quantum computers utilising superposition and entanglement can crack RSA-1024 encryption in approximately 3.58 hours and ECC-256 encryption in about 10.5 hours, requiring around 2,050 and 2,330 logical qubits, respectively, by employing Shor's algorithm, introduced by Peter Shor in 1994 [9]. Furthermore, quantum computers pose a threat to numerous widely used cryptosystems, including elliptic curve DSA (ECDSA), RSA, DSA, and ECDH, through Shor's and Grover's algorithms [10]. While Symmetric Key cryptography is generally more resilient to quantum threats than its asymmetric counterpart, it remains vulnerable due to quantum computers' ability to diminish the effective security of encryption keys significantly. This vulnerability arises from Grover's algorithm, a quantum algorithm developed by Lov Grover in 1996, which can search through unsorted databases far more rapidly than classical methods, locating an item in a database of N entries in approximately $\sqrt{N}$ steps, compared to the classical approach requiring an average of N/2 steps. Grover's algorithm, in particular, threatens symmetric key cryptography by allowing quantum computers to search a billion-item database in about 31,623 steps, compared to 500 million steps for classical algorithms [11].

The "Harvest Now, Decrypt Later" strategy poses a greater threat because attackers can gather encrypted data now, and they are planning to decrypt it in the future when quantum computing technology improves. Given recent advancements, such as IBM's 1,121-qubit processor, IBM Condor, the risk of quantum computers compromising existing cryptographic methods is growing [12].

TABLE I below, based on research by Vlad Gheorghiu and Michele Mosca [13], outlines the qubit requirements and computational resources needed for a quantum computer to break RSA and ECC, underscoring the urgency of transitioning to quantum-resistant cryptographic solutions. Notably, from NIST P-160 to NIST P-521 shown in TABLE I refer to elliptic curve cryptography (ECC) standards established by the National Institute of Standards and Technology (NIST) [14].

TABLE I: Quantum Algorithm Requirements for Various Cryptosystems

| Cryptosystem | Physical Error Rate (pg) | Security Parameter | Quantum Algorithm | Physical Qubits | Logical Qubits | Surface Code Cycles | T Gates Count |
|---|---|---|---|---|---|---|---|
| RSA-1024 | $10^{-3}$ | 80 | Shor's algorithm | $3.01 \times 10^7$ | 2050 | $5.86 \times 10^{13}$ | $3.01 \times 10^{11}$ |
| NIST P-160 | $10^{-3}$ | 80 | Shor's algorithm | $1.81 \times 10^7$ | 1466 | $4.05 \times 10^{13}$ | $2.08 \times 10^{11}$ |
| RSA-2048 | $10^{-3}$ | 112 | Shor's algorithm | $1.72 \times 10^8$ | 4098 | $4.69 \times 10^{14}$ | $2.41 \times 10^{12}$ |
| NIST P-192 | $10^{-3}$ | 96 | Shor's algorithm | $3.37 \times 10^7$ | 1754 | $7.23 \times 10^{13}$ | $3.71 \times 10^{11}$ |
| RSA-3072 | $10^{-3}$ | 128 | Shor's algorithm | $6.41 \times 10^8$ | 6146 | $1.58 \times 10^{15}$ | $8.12 \times 10^{12}$ |
| NIST P-224 | $10^{-3}$ | 112 | Shor's algorithm | $4.91 \times 10^7$ | 2042 | $1.15 \times 10^{14}$ | $5.90 \times 10^{11}$ |
| RSA-4096 | $10^{-3}$ | 128 | Shor's algorithm | $1.18 \times 10^9$ | 8194 | $3.75 \times 10^{15}$ | $1.92 \times 10^{13}$ |
| NIST P-256 | $10^{-3}$ | 128 | Shor's algorithm | $6.67 \times 10^7$ | 2330 | $1.72 \times 10^{14}$ | $8.82 \times 10^{11}$ |
| RSA-7680 | $10^{-3}$ | 192 | Shor's algorithm | $7.70 \times 10^{10}$ | 15362 | $2.64 \times 10^{16}$ | $1.27 \times 10^{14}$ |
| NIST P-384 | $10^{-3}$ | 192 | Shor's algorithm | $2.27 \times 10^8$ | 3484 | $6.17 \times 10^{14}$ | $3.16 \times 10^{12}$ |
| RSA-15360 | $10^{-3}$ | 256 | Shor's algorithm | $4.85 \times 10^{12}$ | 30722 | $2.24 \times 10^{17}$ | $1.01 \times 10^{15}$ |
| NIST P-521 | $10^{-3}$ | 256 | Shor's algorithm | $6.06 \times 10^8$ | 4719 | $1.56 \times 10^{15}$ | $7.98 \times 10^{12}$ |

*B. Quantum Threats to IoT*

Quantum-enhanced attacks exploit the distinct vulnerabilities present in the four layers of the IoT infrastructure-physical, network, perception, and application layers; which impact data integrity, confidentiality, and system functionality. These layers rely on various cryptographic algorithms, which quantum computing can compromise through brute-force decryption, key recovery attacks, and side-channel exploitation. TABLE II presents a detailed overview of quantum attack types across IoT layers, listing the cryptographic algorithms currently in use and description of vulnerabilities against quantum threats.

Internet of Things (IoT) devices also face significant hardware-based security risks. Quantum computing exacerbates these vulnerabilities. Side-channel attacks (SCAs), such as Power Analysis, Electromagnetic Emission Analysis, and Differential Electromagnetic Analysis (DEMA), exploit physical characteristics like power consumption and electromagnetic emissions to extract sensitive data, including cryptographic keys [15]. Fault Injection Attacks (FIAs), specifically Electromagnetic Fault Injection (EMFIs), manipulate environmental parameters (e.g., voltage or clock signals) to induce hardware faults and these faults can grant unauthorised access to critical system functions or reveal confidential information [15]. Additionally, the covert inclusion of Hardware Trojans and Counterfeit Chips, Reverse Engineering during manufacturing introduces further vulnerabilities in the IoT devices' Hardware [15]. Given these attack vectors, robust security measures are crucial for IoT hardware, especially in the face of evolving quantum computing.

Healthcare IoT devices also use standard cryptography algorithms to protect the transmission and storage of sensitive patient data. To guarantee data confidentiality and integrity, widely used algorithms like RSA, AES, and ECC are now in use. However, current conventional encryption techniques used in the Healthcare sector are anticipated to become more vulnerable to breach in the near future due to the development of potent quantum computing capabilities. Hackers could use quantum computer attacks in the future to hack healthcare IoT devices and access, read, and change private healthcare information, including medical records such as patients names, diagnoses, allergies, treatment histories, and so on, jeopardising the confidentiality and integrity of healthcare data. Real-time health data, such as heart rate, blood sugar levels, and even brain activity, will be exposed to the quantum computer assault. Hackers might get control of crucial IoT devices such as pacemakers, insulin pumps, smart inhalers, and others, potentially putting lives at risk.

III. QUANTUM THREATS IN IOT HEALTHCARE

Quantum computing represents a revolutionary approach for processing information, leveraging the principle of quantum mechanics such as superposition and entanglement to solve problems that are intractable for classical computers. It uses the principles of quantum mechanics to carry out intricate calculations that are too difficult for traditional computers in a matter of seconds. Although there are many potential applications for this development, there are also many drawbacks, especially in the field of cybersecurity. With the rapid improvements in quantum technology, the healthcare sector is reaching a turning point when traditional encryption techniques used to protect sensitive patient data might not be adequate.

As healthcare technology continues to evolve, particularly with the advent of quantum computing, the need for robust, quantum-resistant cybersecurity frameworks becomes paramount. Our research focuses on fortifying the security of IoT-enabled healthcare infrastructures to ensure the continued protection and reliability of these essential services in the era of quantum computers.

TABLE II: Quantum Attack Types and Cryptographic Vulnerabilities Across IoT Layers

| IoT Layer | Quantum Attack Types | Attack Description | Cryptographic Algorithms Used |
|---|---|---|---|
| Physical Layer | Quantum tampering, Quantum brute force, MPIA, Attacks based on HHL and QKD | Attacks on hardware components, manipulating or damaging IoT devices, compromising their operation and data integrity. | AES (symmetric encryption), ECC (key exchange), HMAC (integrity verification) |
| Network Layer | Quantum Insert, Quantum key recovery, Quantum man-in-the-middle, Quantum saturation, virus invasion, DoS attack | Attacks targeting data transmission, intercepting, altering, or decrypting communications between IoT devices and central systems. | TLS (RSA/ECDH), WPA2/WPA3, AES-GCM |
| Perception Layer | Quantum jamming & trojan horse, Quantum desynchronising, Quantum DDoS, Timing and Routing Threats, Node Capture | Attacks on IoT sensors and data collection, causing disruptions and unauthorized access to sensitive information. | AES (for encryption), SHA-2 (for integrity), ECC (for key management) |
| Application Layer | Attacks on cloud containers, Blockchain, Quantum hammer, Quantum laser damage, Quantum State Attacks | Compromise of IoT applications, including cloud services and blockchain, leading to data loss and operational failures. | RSA (for digital signatures), AES-256 (for encryption), SHA-256 (for blockchain integrity) |

## A. Mapping the Quantum Attack Surface in a Comprehensive IoT-Based Healthcare Network Design

Fig. 2 shows a comprehensive IoT-based healthcare network that identifies the quantum attack surface in 4 layers (Physical, Network, Perception, and Application), illustrating the flow of data from IoT devices through communication protocols and network infrastructure to data storage solutions and cryptographic assets. Beyond traditional IoT healthcare models, this figure encompasses new technologies like 6G-driven ultra-reliable low-latency communication (URLLC), AI-driven automation in Healthcare 5.0, cyber-physical systems in Industry 5.0, digital twins for real-time patient monitoring, and metaverse-based telemedicine applications. These technologies all pose new security challenges in the quantum era. At the bottom, the figure illustrates how wired and wireless IoT devices, including 6G-enabled healthcare sensors, digital twin interfaces, and metaverse-based remote health applications, transmit real-time patient data to IoT gateways via hotspots, SDN-based routers, network slices, edge nodes, switches, hubs, and local area networks (LANs). This mapping of physical and network layers highlights the evolving quantum attack surface in next-generation healthcare ecosystems, with the corresponding quantum threats to these IoT layers detailed in TABLE II. Secure wireless communication between nodes and gateways requires robust post-quantum encryption due to the growing complexity of eavesdropping techniques [16]. However, the cryptosystems (NIST P-160, P-192, P-224, P-256) frequently used for IoT device encryption are especially susceptible to quantum computing attacks, with the required logical qubits and resources to compromise them detailed in TABLE I. The gateways aggregate and preprocess the data before storing it in on-premises servers or cloud storage solutions. We have identified the perception layer in this process, with the associated quantum attacks on IoT devices outlined in TABLE I. To ensure the availability, integrity, and security of patient data, storage solutions must comply with regulatory standards such as HIPAA, GDPR, and ISO/IEC 27001 [17], particularly as data flows between cloud infrastructures, federated learning models, and blockchain-based medical records in Healthcare 5.0 environments.

The uppermost layer of the figure showcases essential storage services and cryptographic assets, including PKI servers, HSMs, CAs, Authentication Servers, and IAM systems. These components play a pivotal role in storing crucial data, managing encryption keys, digital certificates, and access control, and ensuring data security throughout its lifecycle. Furthermore, the application layer, which encompasses cloud servers, AI-driven healthcare platforms, metaverse-enabled virtual consultations, and cloud-integrated smart hospitals, faces significant quantum security challenges. As illustrated in TABLE I, these connected healthcare ecosystems are susceptible to cryptographic vulnerabilities that quantum adversaries could exploit. Additionally, the figure illustrates the interaction between remote patients using mobile applications and secure VPN connections to transmit health data from devices like pacemakers to the healthcare network. This remote patient case scenario shown in this figure includes 4 IoT layers, and those layers corresponding quantum attacks are shown in TABLE I. This figure also shows the connectivity of healthcare personnel, patient portals, telehealth platforms, phones, remote employees, and remote patients to the Internet, ultimately linking to on-premise and cloud systems. Additionally, the figure highlights the interconnected nature of Healthcare 5.0, where on-premise infrastructure, cloud-based systems, and decentralized IoT nodes interact seamlessly. The integration of smart hospital ecosystems, Industry 5.0 automation, healthcare worker interfaces, patient portals, remote workforce applications, and blockchain-driven medical records presents a broad attack surface for quantum-enabled threats. This comprehensive view underscores the critical significance of a robust framework for migrating to a quantum-safe state to safeguard sensitive patient data against evolving cyber threats, particularly in light of advancements in quantum computing capabilities.

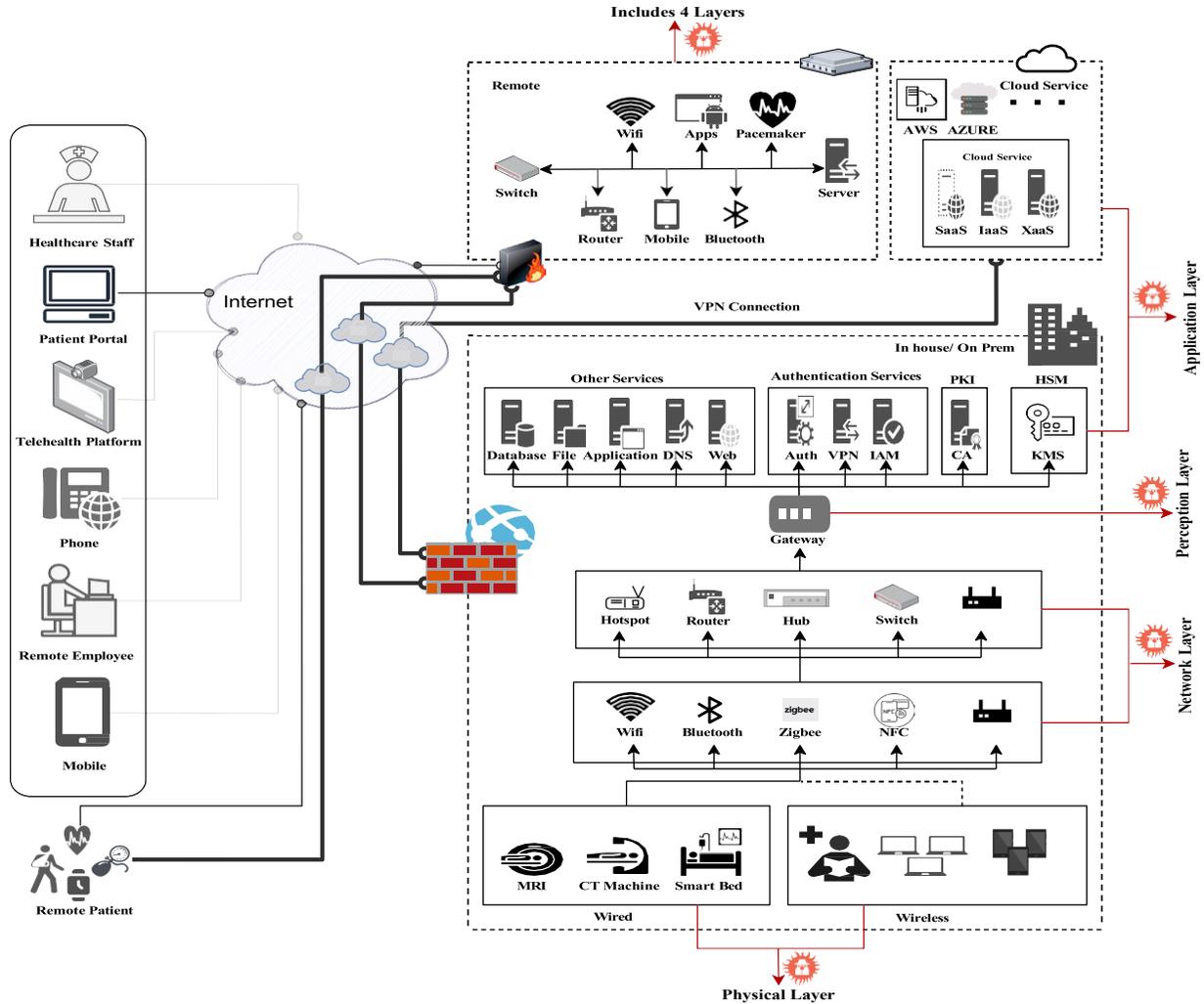

Fig. 2: Visual Representation of Mapping the Quantum Attack Surface in a Comprehensive IoT-Based Healthcare Network

## IV. MIGRATION PLAN FOR IoT-BASED HEALTHCARE SYSTEM

Post-Quantum Cryptography (PQC) is a field focused on designing cryptosystems and protocols that can withstand quantum attacks. It aims to develop new cryptographic methods based on mathematical challenges that are believed to be difficult for quantum computers to solve. The National Institute of Standards and Technology (NIST) is leading an effort to develop and standardise post-quantum cryptographic algorithms resistant to quantum attacks [2]. Given the presence of low-end IoT devices in healthcare systems, integrating PQC can be particularly challenging due to their limited computational resources and processing power. The vulnerabilities in cryptography and IoT hardware to multi-layered quantum threats, it is imperative for IoT-based healthcare systems to promptly transition to a quantum-safe state, following a well-structured and strategic migration plan. Recent efforts by various organisations and researchers have yielded insights into PQC migration strategies across different sectors. Despite the extensive reviews of these strategies, none offer a comprehensive and fully detailed migration framework specifically tailored for IoT-based healthcare systems.

TABLE III presents a comparative evaluation of various Post-Quantum Cryptography (PQC) migration frameworks originating from standard bodies and Govt. drafts, industry and academia, including an overview of their specific migration techniques. The table underscores key gaps in addressing IoT-specific requirements, particularly in contexts where resource limitations and real-time communication are pivotal for achieving secure and effective PQC migration.

In light of these gaps, our proposed framework aims to address these deficiencies by integrating IoT-specific decisions into a phased migration approach. This phased approach is advantageous as it advocates for hybrid cryptography and cryptographic agility, which are well-suited for organisations with complex infrastructures or limited resources, such as IoT-based healthcare systems [24], [25]. By adopting our tailored migration framework, IoT-based healthcare systems can enhance their resilience against the quantum threats illustrated in Fig. 2 and detailed in TABLE II, ensuring both robust security and operational efficiency throughout the migration process. Hybrid cryptography is defined as the use of a post-quantum cryptographic system combined with another public-key cryptographic system (whether post-quantum or traditional) to leverage their strengths, ensuring enhanced security and resilience against both current and future quantum threats while cryptographic agility describes a system or architecture designed to accommodate changes in cryptographic algorithms with minimal impact on functionality, cost, or security, enabling smooth transitions between different cryptographic methods, including the adoption hybrid cryptography [19], [26]. The combination of a hybrid cryptography with cryptographic agility provides significant advantages for complex systems such as IoT-based healthcare. Unlike other frameworks that are available on for IoT or other organisations, our proposed framework specifically addresses the unique challenges of resource-constrained IoT devices in healthcare, enables gradual integration of post-quantum cryptography, allowing for smoother transitions, minimized disruptions ensuring a seamless transition without compromising critical operations.

TABLE III: Comparative Analysis of PQC Migration Frameworks Highlighting IoT-Specific Gaps

| Framework Origin | Framework/ Recommendation [year] | PQC Migration Techniques | Cryptographic Inventory | Backward Compatibility | Interoperability | Quantum Risk Assessment | Dependency Assessment | Ref. |
|---|---|---|---|---|---|---|---|---|
| Standard bodies and Govt. drafts | NIST-SP 1800-38c Preliminary Draft [2023] | Hybrid and Optimised PQC Implementations | Yes (Not IoT-specific) | Yes (Not IoT-specific) | Yes (Not IoT-specific) | Yes (Not IoT-specific) | Not Addressed | [18] |
| | Canadian National Quantum-Readiness [2023] | Phased approach | Yes (Not IoT-specific) | Yes (Not IoT-specific) | Yes (Not IoT-specific) | Yes (Not IoT-specific) | Partially Addressed | [19] |
| | ETSI[2020] | Staged Approach | Yes (Not IoT-specific) | Yes (Not IoT-specific) | Yes (Not IoT-specific) | Yes (Not IoT-specific) | Not Addressed | [20] |
| Industry | IBM Z [2022] | Multi-phase roadmap | Yes (Not IoT-specific) | Not Addressed | Yes (Not IoT-specific) | Yes (Not IoT-specific) | Partially Addressed | [21] |
| | Microsoft [2023] | Multi-phase roadmap | Yes (Not IoT-specific) | Not Addressed | Yes (Not IoT-specific) | Yes (Not IoT-specific) | Partially Addressed | [22] |
| Academia | K. F. Hasan [2024] | Phased approach | Yes (Not IoT-specific) | Yes (Not IoT-specific) | Yes (Not IoT-specific) | Yes (Not IoT-specific) | Addressed | [2] |
| | CARAF [2021] | Phased approach | Yes (Not IoT-specific) | Not Addressed | Not Addressed | Partially Addressed | Not Addressed | [23] |

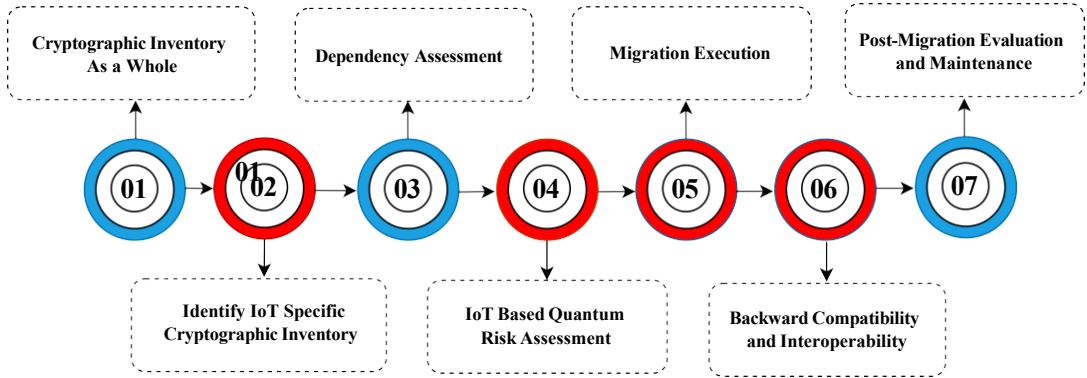

Fig. 3: Proposed migration framework for IoT-based healthcare systems

Fig. 3 presents the proposed framework tailored for IoT-based healthcare systems. This comprehensive framework unfolds across seven distinct phases, each meticulously designed to facilitate a secure and efficient transition to quantum-safe cryptography. The red-shaded area in Fig. 3 highlights the four phases of the framework those are phases 2,4,5 and 6 where IoT-related decisions are addressed.

The first step in obtaining quantum resistance is to create a comprehensive cryptographic inventory for the whole IoT-based healthcare system. The initial phase in this migration framework is to identify all cryptographic assets, such as algorithms, key exchange protocols, hashing techniques, digital signature mechanisms, and libraries used in software and hardware. It sets the groundwork for determining which elements are most sensitive to quantum attacks. Organisations can understand the scope of potential exposure, prioritise systems based on sensitivity and usage, and eliminate outdated or undocumented cryptographic dependencies by mapping out the current cryptographic ecosystem-a critical step in any informed and secure migration strategy.

After completing the overall inventory, the second phase of this migration framework focuses on identifying and cataloging IoT-specific cryptographic artifacts, which are often more limited than those in general-purpose systems. This phase lists all cryptographic activities used in low-end healthcare Internet of Things devices such as wearable monitoring, pacemakers, insulin pumps, smart infusion systems, and others, which often utilise lightweight cryptography and may be vulnerable to quantum assaults. To decide the optimal post-quantum cryptographic algorithms to be deployed during the migration phase, an evaluation is required that evaluates memory restrictions, processor capabilities, energy constraints, and communication protocols. This careful filtering is critical for ensuring the feasibility of PQC adoption without degrading device performance or reliability-two essential factors in healthcare delivery.

The third phase, Dependency Assessment, examines the interdependencies between cryptographic functions and other system components such as firmware, cloud services, EHR databases, and third-party interfaces. In healthcare IoT, devices frequently function in tightly connected ecosystems, where a failure in one cryptographic service might spread across numerous levels. So, in this case, downtime might be fatal. This step ensures that these dependencies are thoroughly understood so that PQC updates may be coordinated across interrelated components without causing cascade interruptions.

The fourth phase, IoT-Based Quantum Risk Assessment, assesses and prioritises quantum-specific hazards in IoT settings, including vulnerabilities and the impact of quantum attacks on IoT devices and cryptographic systems. Not all data or devices pose the same danger or have the same shelf life. This step determines which IoT-based cryptographic assets require rapid migration by assessing their exposure, sensitivity, and retention timelines. To prioritise mitigation techniques and improve security against quantum risks, organisations can apply the formula Migration Time + Shelf-life Time > Threat Timeline [19]. For example, a pacemaker's firmware or a patient's long-term medical record may need to be safeguarded much beyond the anticipated advent of practical quantum computers. By focusing first on systems whose confidentiality must be preserved for decades, this phase ensures that the most impactful components of the healthcare infrastructure receive quantum protection early in the migration lifecycle.

The fifth phase, Migration Execution, focuses on the implementation of Post-Quantum Cryptography (PQC) specifically designed for healthcare IoT environments. This involves selecting and rigorously testing PQC algorithms for IoT devices. Rigorous testing is conducted to evaluate performance on constrained IoT devices, ensuring minimal disruption to real-time healthcare operations. At this phase, selected PQC algorithms are deployed across prioritised systems. A hybrid cryptography approach should be employed, starting with the most critical systems to minimise disruption and preserve data integrity during the transition. Since immediate and complete replacement is impractical in healthcare environments, hybrid cryptography allows secure coexistence during transition. Additionally, incorporating cryptographic agility is crucial, allowing for future-proofing by enabling flexible updates and adaptations to new quantum- resistant technologies as they become available. This phase directly contributes to quantum resistance by embedding robust, upgrade-ready cryptography into the system's core.

IoT-based healthcare systems frequently include a mix of legacy and contemporary equipment. The sixth phase, Backward Compatibility, permits older IoT devices to stay operational during the migration while simultaneously reviewing Interoperability to verify that new PQC protocols connect easily with current systems, hence ensuring overall system coherence. Maintaining backward compatibility is critical for progressive adoption and decreases the possibility of service gaps during migration, protecting operational integrity while transitioning to a completely quantum-resistant infrastructure.

The final phase, Post-Migration Evaluation and Maintenance, guarantees that the quantum-safe transition is sustainable in the long run. It entails continuous auditing, performance benchmarking, cryptographic patching, standard compliance tests, and the use of monitoring tools to detect developing vulnerabilities. This ongoing repair cycle is critical in healthcare systems, where lives depend on uninterrupted service. Post-migration evaluation also gives input for future migrations and initiates an iterative improvement cycle, guaranteeing that the healthcare IoT ecosystem is resilient to both present and future quantum threats.

This migration framework is specifically designed to preserve both the security and operational integrity of IoT devices within healthcare environments-where any disruption could have life-threatening consequences. This framework is better than the existing frameworks from standard bodies, academia and industry players as shown in TABLE III, which typically present generalised strategies or overlook the limitations of resource-constrained IoT devices, where our framework offers a detailed, phased roadmap explicitly tailored for IoT-based healthcare infrastructures. A key differentiator of our framework is it incorporates IoT-layer-specific risk assessments, phased cryptographic upgrades, hybrid cryptography adoption, and cryptographic agility to support long-term adaptability. The 7 phases of our framework are carefully structured to ensure that PQC adoption is gradual, risk-aware, and minimally disruptive to existing healthcare operations with the red shaded stages in Fig. 3 highlight phases that directly involve IoT-specific decision-making and are critical for aligning cryptographic upgrades with device constraints and network protocols which is better than the existing frameworks.

Although the proposed framework defines a complete plan for transitioning healthcare IoT systems to post-quantum cryptography (PQC), a number of practical issues remain. PQC integration is particularly problematic in low-power IoT devices because to limited memory, computing capability, and energy restrictions. Maintaining interoperability and backward compatibility with outdated systems increases complexity, particularly in diverse healthcare contexts. Future work will include establishing the framework in real-world healthcare networks and simulation-based testbeds to determine its scalability, performance overhead, and flexibility across several device ecosystems. Additional research will focus on refining PQC algorithms for limited contexts and collaborating with standards organisations to promote the widespread adoption, interoperability, and long-term durability of quantum-secure healthcare infrastructures.

## V. Case Study: Securing Remote Glucose Monitoring for Diabetics

Continuous Glucose Monitors (CGMs) have transformed diabetes management by providing real-time glucose readings without requiring frequent finger pricks [1]. These devices wirelessly transmit data to smartphones, hospital servers, and cloud-based healthcare platforms, making them vulnerable to quantum-enabled cyber threats. Existing encryption schemes such as ECC-P256 and AES-128 face imminent risks from Shor's and Grover's algorithms, allowing adversaries to decrypt, intercept, and manipulate glucose readings. Additionally, CGMs operate under severe resource constraints (32-64 KB RAM, 1-2 MB ROM, 16-32 MHz clock speed), exposing them to side-channel attacks, firmware exploits, and unauthorized access [1]. These challenges necessitate a structured quantum risk assessment-mapping threats such as man-in-the-middle (MitM) attacks, key recovery vulnerabilities, and quantum-assisted decryption, which our seven-phase PQC migration framework Fig. 3 systematically addresses.

To mitigate these vulnerabilities, our framework integrates PQC algorithms tailored to each IoT layer. At the physical layer, Kyber (for key exchange) and Dilithium (for authentication) secure device pairing between CGMs and Bluetooth/Wi-Fi-connected devices. For the network layer, where glucose data is transmitted via Bluetooth, Wi-Fi, and 6G, FrodoKEM combined with post-quantum TLS 1.3 ensures end-to-end encryption against quantum-assisted traffic decryption. The perception layer, which includes CGMs and IoT biosensors, requires SPHINCS+ for secure firmware updates and hybrid PQC-enabled VPNs to prevent malicious data injection. Finally, at the application layer, where patient glucose data is processed and stored, PQC-secured cloud storage and blockchain-based medical data integrity solutions protect against quantum decryption attacks on electronic health records (EHRs). Through this layered PQC integration, aligned with our seven-phase migration approach, CGM systems transition seamlessly into quantum-safe healthcare environments, ensuring long-term security and patient safety while maintaining backward compatibility with legacy healthcare infrastructures.

## VI. Conclusion

This paper focusses on the urgent need to address the vulnerabilities of healthcare IoT systems in the face of rapidly advancing quantum computing. Quantum attacks, as analysed in this paper, pose a significant threat to the cryptographic algorithms commonly employed in these systems, which has the potential to jeopardise patient data and critical healthcare operations. To mitigate these risks, this article introduces a comprehensive migration framework for transitioning healthcare IoT to quantum-safe cryptography. This framework, based on a phased approach offers a practical roadmap for ensuring the long-term security and resilience of healthcare IoT systems in the era of quantum computing. As a future research direction, we aim to implement and evaluate the proposed framework in simulated and real-world healthcare IoT settings. This includes benchmarking various PQC algorithms based on performance metrics such as Handshaking time, latency, memory usage, and energy efficiency to identify the most suitable options for resource-constrained medical devices. Future work will also involve developing automated tools for cryptographic inventory and quantum risk analysis, as well as enhancing cryptographic agility in dynamic healthcare networks. Establishing standardised testbeds will be crucial for validating interoperability and system resilience across diverse clinical scenarios.